\documentclass[twocolumn,runningheads,natbib]{svjour2}

\smartqed  % flush right qed marks, e.g. at end of proof
\usepackage[]{graphicx}                      %wichtig z.B. auch fuer bunte Sachen 
\usepackage{threeparttable}
\usepackage{amsmath}
\usepackage[latin1]{inputenc} % umlaute etc.
\usepackage{latexsym} % einige sonderzeichen
\newcommand\aj{{AJ}}% 
\newcommand\araa{{ARA\&A}}% 
\newcommand\apj{{ApJ}}% 
\newcommand\apjl{{ApJ}}% 
\newcommand\apss{{Ap\&SS}}% 
\newcommand\aap{{A\&A}}% 
\newcommand\aaps{{A\&AS}}% 
\newcommand\mnras{{MNRAS}}% 
\newcommand\gca{{Geochim.~Cosmochim.~Acta}}% 
\newcommand\HII{H\,{\sc ii} }
\newcommand\HI{H\,{\sc i} }

\journalname{Astrophysics and Space Science}

\begin{document}

\title{The Magnificent Seven in the dusty prairie
%Insert your title here%\thanks{Grants or other notes
}
\subtitle{The role of interstellar absorption on the observed neutron star population
% of thermal isolated neutron stars 
}


\author{B. Posselt \and S.~B. Popov \and F. Haberl \and  J. Tr\"{u}mper  \and R. Turolla \and  R.~Neuh\"{a}user
%        Second Author %etc.
}


\institute{B. Posselt \at
              Max-Planck-Institut f\"{u}r extraterrestrische Physik /\\                        Universit\"{a}tssternwarte Jena \\
	      Giessenbachstrasse,  85748 Garching, Germany\\
              Tel.: +49-3641-947538\\
%              Fax: +123-45-678910\\
              \email{posselt@mpe.mpg.de}           %  \\
           \and
S.B Popov \at
              Sternberg Astronomical Institute \\
              119992 Russia, Moscow, Universitetski pr. 13\\  
              Tel.: +7-495-9395006\\   
%              Fax: +7-495-9328841\\    
              \email{polar@sai.msu.ru}           %  \\
%             second address
}

\date{Received: date / Accepted: date}

\maketitle

\begin{abstract}
The Magnificent Seven have all been discovered by their exceptional soft
X-ray spectra and high ratios of X-ray to optical flux. They all are
considered to be nearby sources. 
Searching for similar objects with larger
distances, one expects larger interstellar absorption resulting in harder
X-ray counterparts. Current interstellar absorption treatment depends on
chosen abundances and scattering cross-sections of the elements as well as
on the 3D distribution of the interstellar medium. 
After a discussion of these factors we 
use the comprehensive 3D measurements of the Local Bubble
by \cite{Lall03} to construct two simple models of the 3D
distribution of the hydrogen column density. We test these models 
by using a set of soft X-ray sources with known distances. Finally, we
discuss possible applications for distance estimations and population
synthesis studies.
%We will discuss this effect and the importance of considering hardness ratios to find new XDINSs. Furthermore we will present our results on the search for new XDINS candidates using ROSAT PSPC / HRI and XMM pointings in combination with the relatively deep optical Sloan Digital Sky Survey.
%Insert your abstract here. Include up to five keywords.
%\PACS{First \and Second \and More}
\end{abstract}
\vspace{-0.7cm}
\section{Introduction}
\label{intro}
The Magnificent Seven (M7), as the ROSAT-discovered X-ray thermal isolated neutron stars are sometimes dubbed, are exceptional because of their soft blackbody-like spectra without non-thermal components; for reviews see e.g. Haberl 2006 (this volume) or \cite{Haberl2004b}. 
One of them has the second-best blackbody spectrum after the cosmic background radiation and this lead to one of the best neutron star radius estimates known \citep{Trumper04}. Thought to be nearby, the M7 represent nearly half of the local young neutron star population.
Despite several searches for new candidates of this radio-quiet neutron star class (e.g. \citealt{Marcel2006,Chieregato2005,Rutledge2003}), no new M7-like object has been confirmed up to now. One aggravating circumstance these searches face is the hardening of the X-ray spectrum due to the interstellar absorption. 
Caused mainly by photo-electric absorption of photons by heavy elements, the interstellar absorption is usually described by the equivalent hydrogen column density assuming certain chemical abundances. 
Highly important is the clumpiness of the interstellar medium  (ISM) causing (un)favourable lines of sights to search for new M7-like objects. The knowledge about the local ISM structure is also important in connection with the long-debated issue of  isolated neutron stars accreting from the ISM.
%Your text comes here. Separate text sections with
\vspace{-0.7cm}
\subsection{Abundances and cross-sections}
\begin{figure}
\centering
  \includegraphics[angle=270,width=0.5\textwidth]{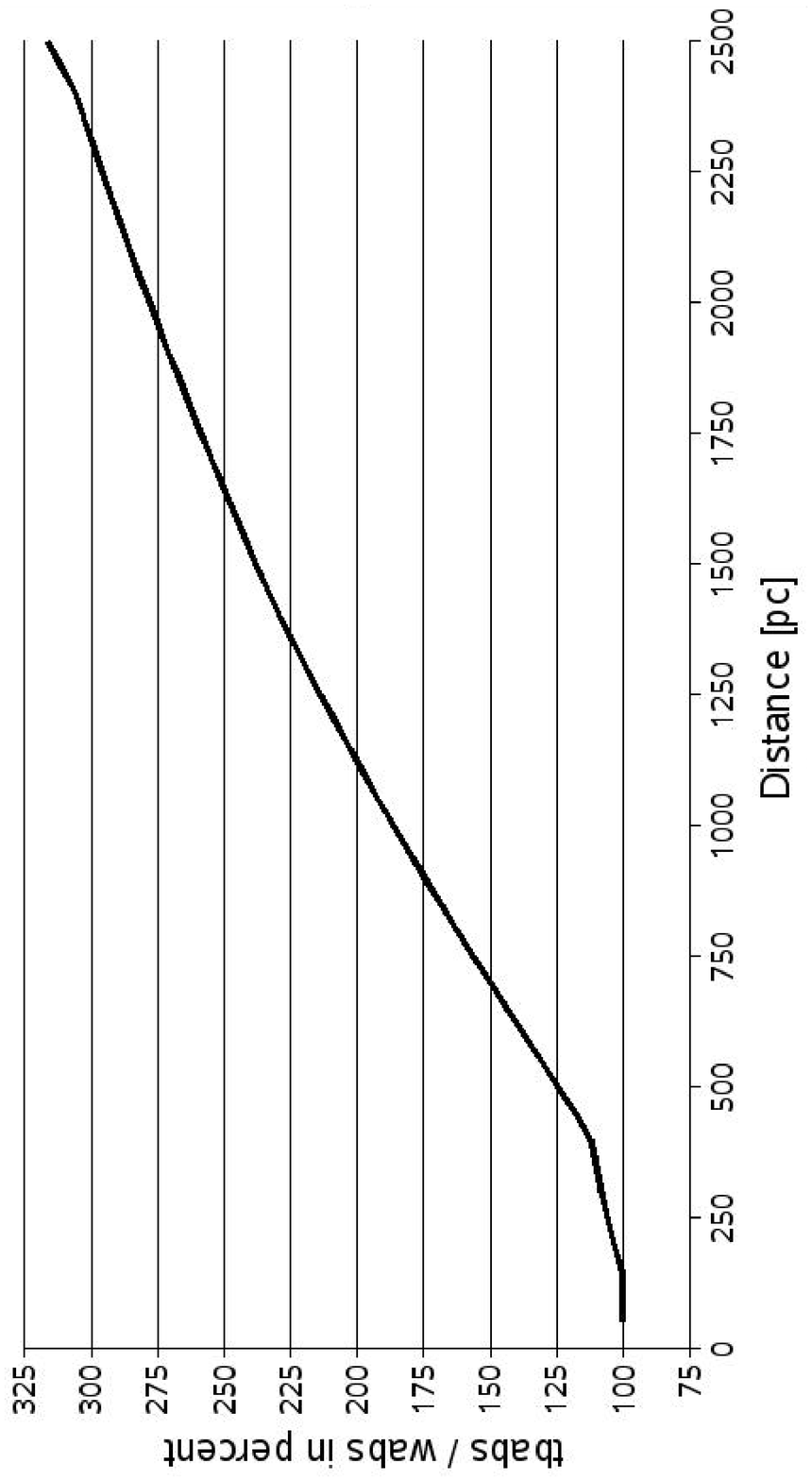}
 % width=0.75\textwidth]{diffWabsTbabs.ps}
\caption{{\bf{The effect of different abundances and cross-sections on the absorption towards soft X-ray sources}} \protect\\
% faint sources%}} \protect\\
For the same neutron star (blackbody of 90~eV, 11.67 km radius, and 1.48 Solar masses) the ratio of the X-ray fluxes (as would be expected to be observed by ROSAT) resulting from the two $XSPEC$ routines '$tbabs$' and '$wabs$' are calculated for different distances. 
The galactic coordinate $x$ varies from 50 to 2500 ~pc (direction: from the Sun towards the galactic center), $y$ is always 0, $z$ is 25 pc north .
For the ISM distribution, we applied the same as used by \cite{Popov2000}, which is described in more detail below.
The effect of less absorption, resulting from the use of the newer routine 'tbabs', is stronger at larger column densities, hence distances. }
\label{fig:abun}       % Give a unique label
\vspace{-0.3cm}
\end{figure}

\vspace{-0.3cm}
\label{abundances}
Most absorption models today presume element abundances independent of the line of sight. It was outlined by \cite{Wilms2000} that the total gas plus dust ISM abundances are lower than the local -- Solar -- abundances and the general ISM abundance uncertainties are still in the order of 0.1~dex because the measurements are very difficult.
% for the total gas plus dust ISM abundances
One can choose between several different abundance tables implemented e.g. in $XSPEC$ for analysing X-ray spectra. We note here only two examples commonly used -- the abundance tables by  \citet{Anders1989} ('angr' in $XSPEC$) and by \cite{Wilms2000} ('wilm' in $XSPEC$) considering newer measurements by e.g. \citet{Snow1996,Cardelli1996,Meyer97,Meyer98}.\\
The individual abundances and photoionization cross-sections of the elements and their ions as well as dust grain properties (e.g. sizes, shapes) influence the total photo-electric absorption of X-rays. Again, we mention here only two examplary works used very often for analysis of X-ray spectra. 
\cite{MM83} did a polynomial fit of the effective absorption cross-section per hydrogen atom based on measured atomic absorption cross-sections by \cite{Henke82} and abundances mostly from \cite{Anders82}. With the exception of oxygen, \cite{MM83} assume the elements to be either entirely in the gas phase or completely depleted into dust grains.  
The recent improved absorption treatment by \cite{Wilms2000} takes into account newer element abundances as noted above and more recent photoionization cross-section calculations by e.g. \cite{Verner93,Verner95}.
\cite{Wilms2000} additionally include an improved molecular cross-section for $H_2$.
%by using a measured photoabsorption cross section of $\approx 2.85$ (\cite{Yan1998}) instead of 2 for $H_2$.  
Besides this updated database \cite{Wilms2000} considered furthermore a simple spherical composite dust grain model in an MRN distribution \citep{MathisRN77}. 
%The composition is represented by actually measured depletion factors, the ratio between the gas phase abundance and the total abundance of an element.
\\
Both models are implemented in $XSPEC$ and can be used via the routines '$wabs$' \citep{MM83} or '$tbabs$' \citep{Wilms2000}.
The various absorption descriptions can lead to different results (see Fig.~\ref{fig:abun}) or expectations of e.g. observable objects. The differences caused by the diverse treatments show our incomplete knowledge about the interstellar absorption and should be kept in mind especially when interpreting soft X-ray spectra.   
\vspace{-0.6cm}
\subsection{The inhomogeneously distributed ISM}
\label{ISMintro}
%For radio pulsars the absorption of radio-pulses is taken into account in the term of the dispersion measure caused by the electron density thought to scale with the distance. The X-ray thermal isolated neutron stars however are radio-quiet with only two claims of weak radio-detections \citep{Malofeev2006,Malofeev2005} still to be confirmed.  
\begin{figure}
\centering
  \includegraphics[angle=0,width=0.35\textwidth]{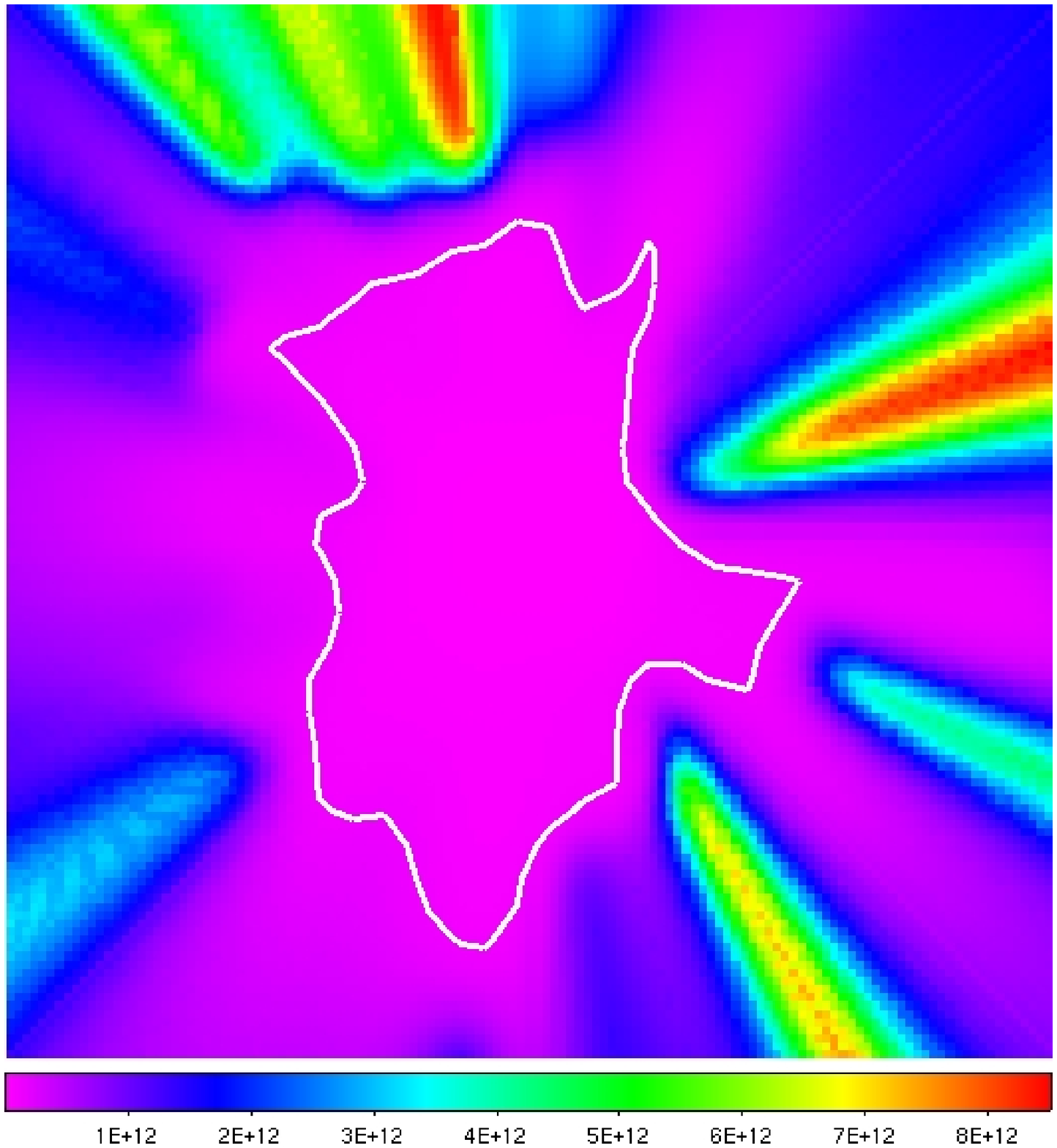}
 % width=0.75\textwidth]{diffWabsTbabs.ps}
\caption{{\bf{The limits of using the \cite{Ferlet1985} formula }} \protect\\
The sodium column density as derived from the density data by \cite{Lall03} is shown in galactic cartesian coordinates at z=0; x is towards the galactic center which is at the right, y towards l=90$^{\circ}$ is pointing up. The image spans 500~pc times 500~pc. The area within the contour  has  low column densities ($\log$ N$(Na I) <11$) for which \cite{Welty1994} doubted whether the \cite{Ferlet1985} formula is applicable. This region of small sodium column densities  is largest at z=0 and becomes smaller for lower or larger z.
}
\label{fig:lal}       % Give a unique label
\vspace{-0.2cm}
\end{figure}
In X-ray astronomy with low spectroscopic resolution the chosen abundances and cross-sections are used to derive the hydrogen column density responsible for the instellar absorption.  
Cold and warm H-atoms (\HI) as well as hydrogen molecules and ionized hydrogen (\HII) contribute to the overall amount of hydrogen along a line of sight. These different components of the hydrogen column density are estimated by different measurement methods, e.g. various extinction measurements as summarized by \cite{Knude2002} or radio 21-cm observations of hydrogen as the extensive study by \cite{Dickey90} and more recently by \cite{Kalberla2005}. A wealth of theoretical models has been applied to these observations. However, most of them are only 2D with few more recent exceptions
%aiming for an description of the galactic spiral structure  
(see e.g.  \citealt{Drimmel2003}, \citealt{Amores2005} or \citealt{Marshall2006astroph} for 3D models).
The overall result is a highly inhomogeneously distributed interstellar medium within the Milky Way, present in  bubbles with loops or dense rims, and shaped by molecular clouds, supernova explosions  and stellar activity (for an illustrated overview see \citealt{Henbest1994}).
The ISM distribution is best known in the close Solar neighbourhood, especially in the Local Bubble (e.g. \citealt{breit1998}). The most detailed view is provided by the study of \cite{Lall03}, who measured Na {\sc{i}} absorption towards 1005 sightlines with Hipparcos distances up to 350 pc from the Sun; for more details on the measurements  and the applied density inversion method see \cite{Lall03}, the precursor paper \cite{Sfeir99}, and  \cite{Vergely01}.
Na{\sc{i}} is regarded as good tracer of the total amount of the neutral interstellar gas and the sodium column density is thought to be convertable into a hydrogen column density (e.g. \cite{Ferlet1985}, but see also discussion in sec.~\ref{sec:models}).
For more distant regions the determination of a 3D distribution of the ISM is hampered mainly by the unknown or imprecise distances of the measured objects and the lack of a well-sampled grid of measurement points. This is true for all extinction-related methods to our knowledge and results in much more  coarse sampling 
%and uncertainties by averaging
compared to the Solar neighbourhood. 
The results obtained by the various methods  can differ for the same regions and a combination can be difficult. 
These difficulties result partly from the different observed astrophysical phenomena like measuring the dust emission or stellar magnitudes. They also result from different sets of stellar absolute magnitude calibrations and the underlying assumptions which can deviate from each other as interstellar extinction determination is an iterative process coupled with the identification of the statistical properties of stars \citep{Hakkila97}. 
\begin{figure}
\centering
  \includegraphics[angle=0,width=0.5\textwidth]{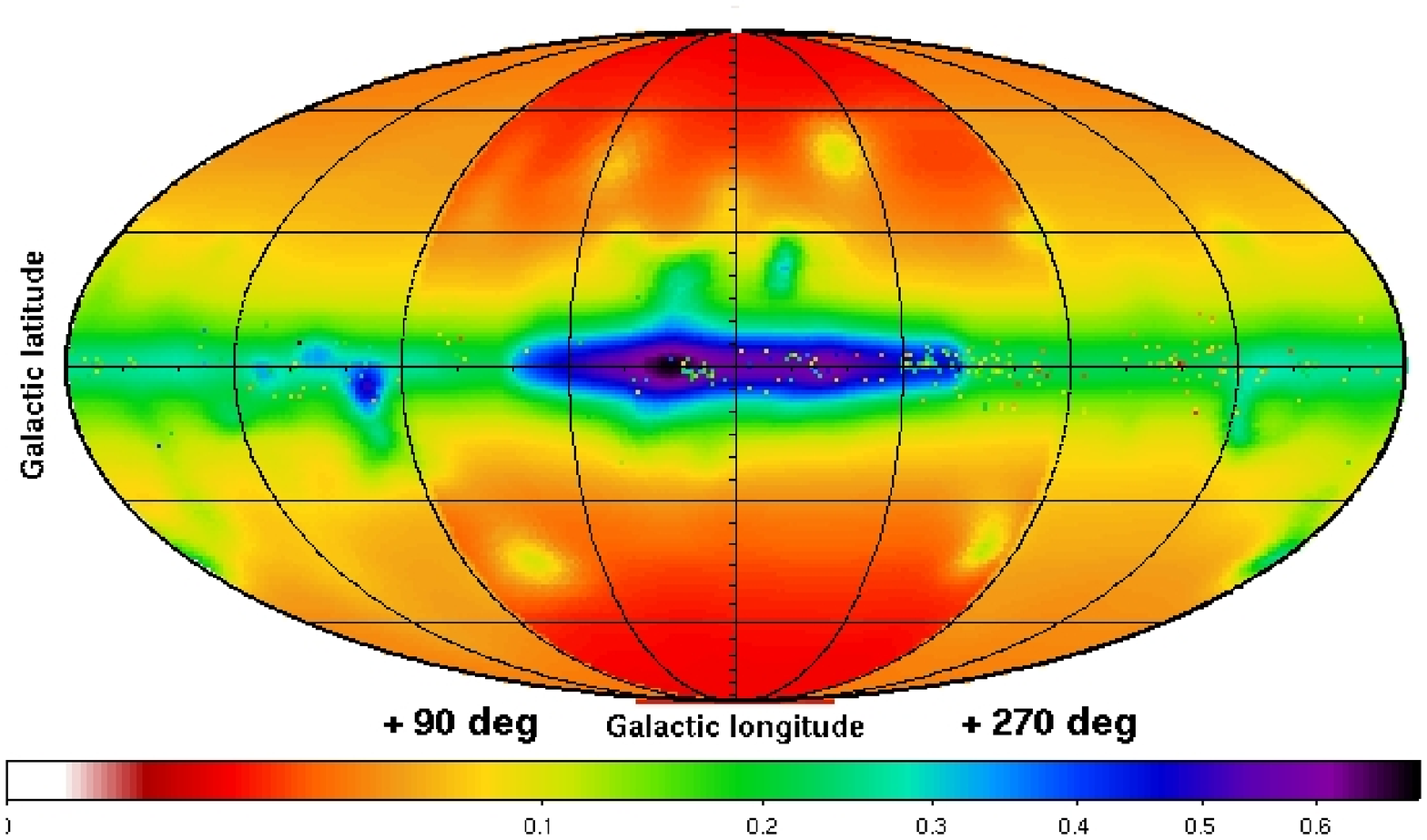}
\caption{{\bf The N(H) predicted by the analytical model at 1~kpc}\protect\\
Positive galactic latitudes are up and $l=0^{\circ}$ is in the center. The N(H)is in units of $10^{22}$cm$^{-2}$.
For description of the model and discussion see text.}
\label{fig:ana}       % Give a unique label
\vspace{-0.2cm}
\end{figure}

\vspace{-0.2cm}
\section{Two simple 3D distribution model cubes of the absorbing ISM}
\label{sec:models}
Considering the absorption in the Solar neighbourhood, the result by \cite{Lall03} is a good starting point representing the currently best database of the local ISM distribution. R.~Lallement kindly provided us with the Na{\sc{i}} density cube derived using the inversion method developed by \cite{Vergely01}. Due to the smoothing length of 25~pc applied in this method by \cite{Lall03}, structures smaller than 25~pc cannot be resolved. We note further that even if measurements go up to Hipparcos distances of 350~pc from the Sun, the sampling becomes coarser at larger distances. Thus the sodium density cube has a span of only 250~pc and starting from 200~pc one has to be careful dealing eventually with the a priori density information applied in the inversion method (R.~Lallement, personal communication).
From the Na{\sc{i}} density cube we calculate the column density for the same grid (sampling 3.9~pc). This is then converted to \HI column density applying the formula by \cite{Ferlet1985}.

It has to be mentioned that there are on-going discussions about how well the sodium D-line absorption actually traces \HI. The  correlation found between the \HI and Na column densities derived from Na-D absorption lines by \cite{Ferlet1985} was doubted by \cite{Welty1994}, especially for low column densities ($\log$ N$(Na I) <11$). The region of the Lallement cube that should not be used because it has such low column densities is  indicated in Fig.~\ref{fig:lal}. 
\begin{figure}
\centering
  \includegraphics[angle=0,width=0.5\textwidth]{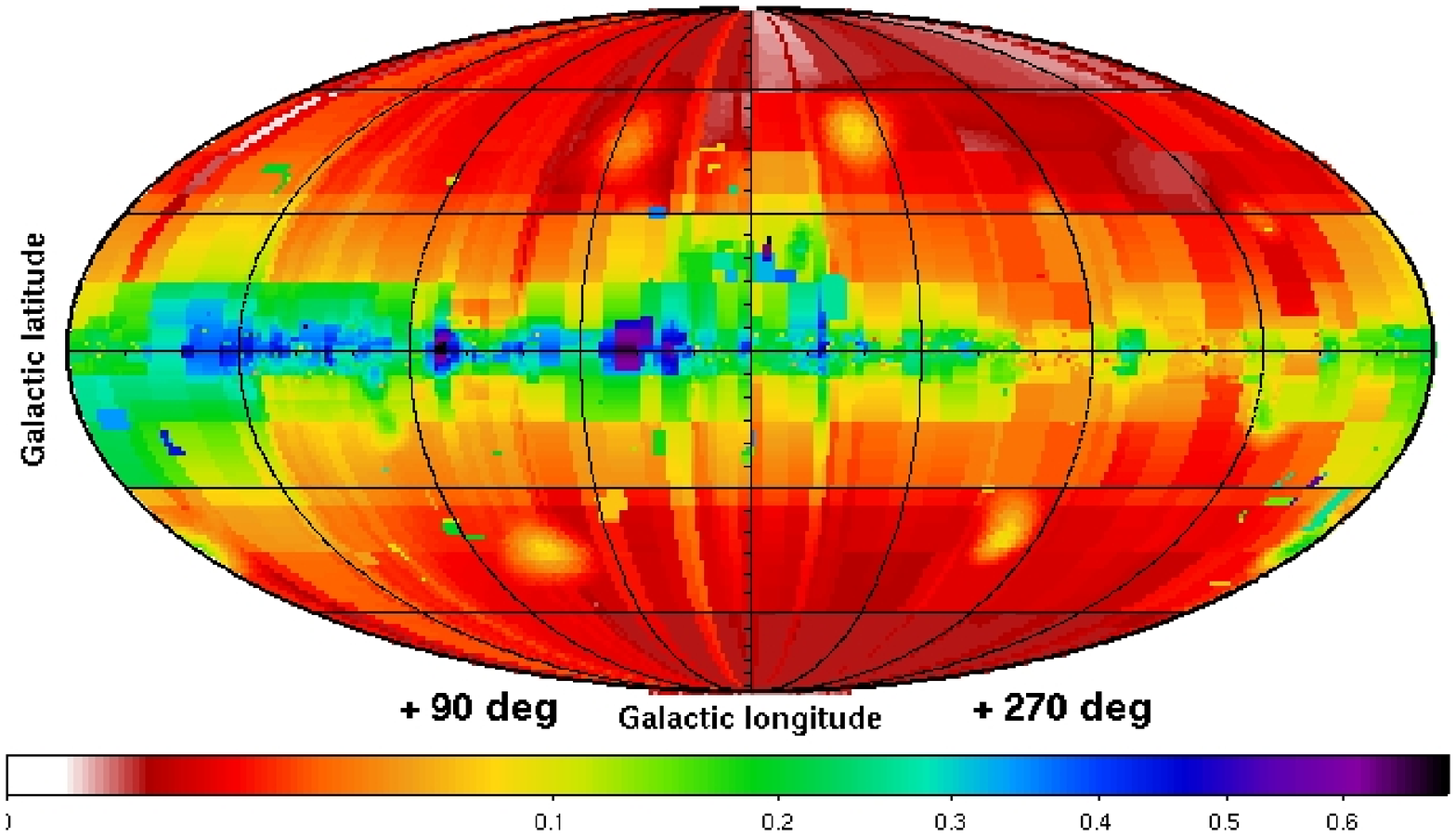}
\caption{{\bf The N(H) predicted by the extinction model based on \citet{Hakkila97} at 1~kpc} \protect\\
Positive galactic latitudes are up and $l=0^{\circ}$ is in the center. The N(H) scale is in unities of $10^{22}$cm$^{-2}$.
For description of the model and discussion see text.}
\label{fig:ext}       % Give a unique label
\vspace{-0.2cm}
\end{figure}

However, the lowest known X-ray--measured hydrogen column density for one of the M7 converts to a sodium column density of $\log$ N(Na{\sc{i}}) $> 11.5$ (applying the formula by \citealt{Ferlet1985}). Therefore this low-column density uncertainty is not important considering the M7 and negligible for the intended population synthesis with neutron stars (NSs) at usually larger distances.\\
\cite{Vergely01} could not always find a correlation of the Na{\sc{i}} density with the \HI density. They concluded that the correlation is rather weak due to non-constant population ratios for interstellar medium species, thus different abundances.  \cite{Lall03} noted the size of the Local Bubble revealed by \HI is smaller than when derived by Na{\sc{i}}. This was explained by the first ionization stage of Na{\sc{i}} being below that of \HI, resulting in a longer neutral phase of \HI \citep{Lall03}. 
Recently, \cite{Hunter2006} presented new ultraviolet Na{\sc{i}} observations towards 74 O- and B-type stars in the galactic disk. Where possible they also derived column densities of the Na{\sc{i}} D-lines.
Then they compared the correlation between the Na{\sc{i}}  and \HI column densities. While there was an excellent agreement with \cite{Ferlet1985} for the D-lines, this was not the case for their ultraviolet absorption lines. \cite{Hunter2006} found a significant offset for the correlation they got from the Na{\sc{i}} D-lines compared to the relation derived from the UV transitions,  even after accounting for saturation effects in the D-lines. They argue that this offset is due to an underestimation of approximately 30 $\%$  of N(H) in the case of Na{\sc{i}} D-lines observations. 
%We consider these extra 30~$\%$ in our discussion  XXX.
Recapitulating, one has to be aware of the still not completely solved problems concerning the conversion of Na{\sc{i}} to \HI column densities when it comes to interpreting individual results. However, we regard the measurements by \cite{Lall03} and the conversion to a hydrogen column density according to \cite{Ferlet1985} as the best possibility at hand to take into account the local inhomogenity in the ISM for X-ray astronomy with low spectral resolution.

\begin{figure}
\centering
  \includegraphics[angle=0,width=0.5\textwidth]{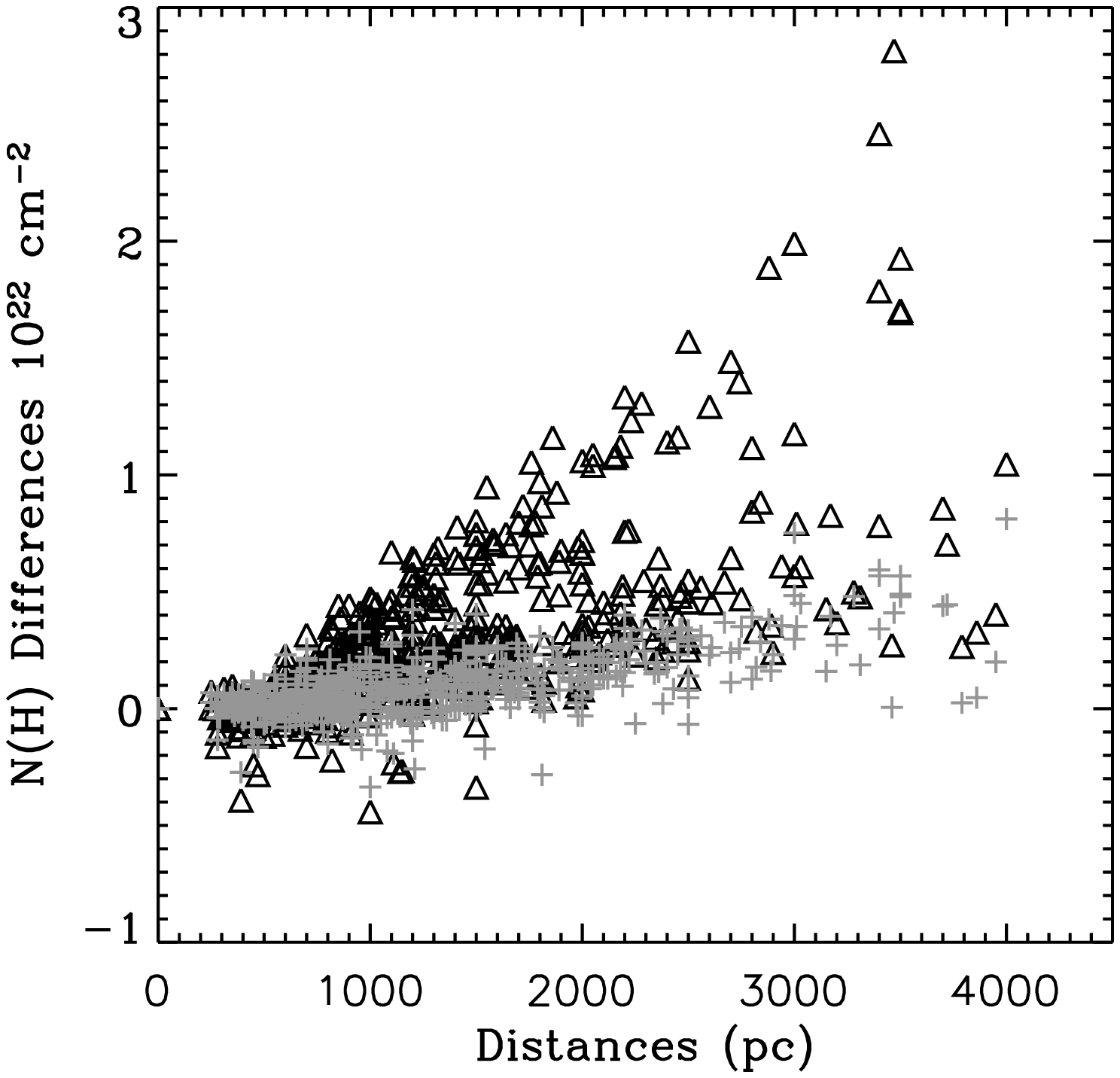}
\caption{{\bf N(H) deviations at different distances\protect\\} Shown are the differences between the model N(H) at the distances of the open cluster sample and the N(H) obtained directly from the reddening measurements of this sample. The results of the extinction model based on \cite{Hakkila97} are plotted as grey crosses, those of the analytical model as black triangles. While the scatter at low distances is relatively small, the analytical model has a larger spread at larger distances.
}
\label{fig:Pisk1}       % Give a unique labe
\vspace{-0.6cm}
\end{figure}

At larger distances we consider two different models -- one is based completely on extinction measurements and another one is described by analytical formulae.  \cite{Hakkila97} put several extinction studies carefully together in an easily accessible routine. All studies have been modified to statistically account for unsampled regions.
%Therefore, the actual measurement points are not easy to identify or extract.
Additionally, a correction method for the systematic underestimation between 1 and 5 kpc was developed. Errors were provided individually for each considered main survey and for their mean. The often large mean error values illustrate the disagreement among  individual observations. 
The model by \cite{Hakkila97} is a large scale model, capable of identifying e.g. molecular clouds at intermediate distances, but not sensitve to extinction variations of less than 1~degree. It is well suited for statistical studies.\\
The analytical model we apply is described by \cite{Popov2000}. It is based on the formulae by \citep{Zane95, Dickey90, deBoer91}. \cite{Popov2000} included additionally a galactocentric radius dependency for the number density of atomic and molecular hydrogen as estimated by \cite{Bochkarev92}. The Local Bubble was taken into account by \cite{Popov2000} as a sphere of 140~pc radius having a constant low density of 0.1 particles cm$^{-3}$. 
As noted above, there are other more recent theoretical 3D extinction models. However, the analytical $H$ density model by \cite{Popov2000} seems to be at least as good as e.g. the underlying model used by \cite{Amores2005}. The model of \cite{Drimmel2003} cannot be applied to distances less than a few hundred parsecs in the Solar neighbourhood and the  data of \cite{Marshall2006astroph} has a coarse sampling with distance bins of 1~kpc.\\
Both described models are applied only for distances larger than 230~pc in case they provide an N(H) larger than derived from the sodium cube at 230~pc. Otherwise the hydrogen column density at 230~pc is taken since the column density can only increase with distance.
We use spherical coordinates -- the galactic cordinates $l$, $b$ and distance $d$. The nominal sampling is one degree in $l$ and $b$, and 10~pc for the distance. This is technically motivated and does not represent the actually achieved accuracy. Distances are covered up to 4500~pc. 
%\emph{good results due to spherical coordinates starting from 120~pc, build in absorption routine warning!}

\vspace{-0.4cm}
\subsection{Testing the models and further improvements}
\begin{figure}
\centering
  \includegraphics[angle=0,width=0.5\textwidth]{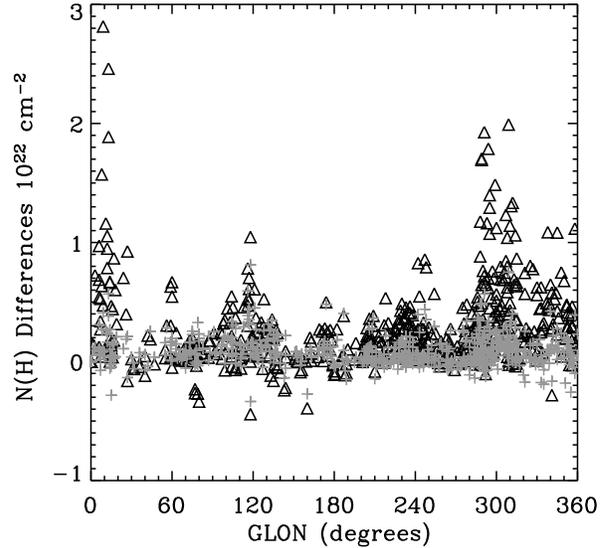}
\caption{{\bf N(H) deviations at different galactic longitudes}
The symbols for the models are the same as in Fig.~\ref{fig:Pisk1}.
For both models a larger scatter is visible at GLON$= 0^{\circ}$,$120^{\circ}$, and $300^{\circ}$ }
\label{fig:Pisk2}       % Give a unique label
\vspace{-0.5cm}
\end{figure}

To test our models we first consider a relatively large number of test objects with good distances and extinction measurements, not necessarily determined from X-rays. Then we proceed to a few of the rare neutron stars with well known distances, having also small error bars for the absorbing N(H) derived by X-ray observations.\\
As \cite{Hakkila97} note, different studies do not agree precisely with each other due to the various applied methods or objects studied. The latter are usually very coarsely scattered and measurement errors can influence the extinction values of a large region. One possibility to overcome this problem is investigating open clusters.
An open cluster has the advantage that one can measure a number of stars at approximately the same distance, which is usually well known. Therefore, open clusters are a good choice in aiming to minimize the extinction measurement errors along one line of sight. Usually one measures the reddening $E(B-V)$.
We concentrate here on the recent comprehensive reddening measurements for 650 open clusters by \cite{Piskunov2006}, representing a complete sample up to distances of about 850~pc from the Sun. Details about this open cluster study can be found in \cite{Piskunov2006, Khar2005a, Khar2005b}.
To convert the reddening $E(B-V)$ we apply the formula by \cite{Paresce84}: $
N(H)=5.5 \times 10^{21} E(B-V)  \left[\rm{cm}^{-2}\right]$.
This is quite similar to the slope found in the X-ray study for the extinction $A_V$ by \cite{Predehl95}: $ 
N(H)=1.79 \pm 0.03 \times 10^{21} A_V -0.41 \left[\rm{cm}^{-2} \right]
$
if $ A_V=3.1 E(B-V)$. This is noted here because there is a deviation of the correlation by \cite{Predehl95} between $A_V$, indicator of the dust, and $N(H)$, indicator for cold gas and dust, at low distances. This is probably due to a significantly lower amount of dust in the Local Bubble, as also found by \cite{Predehl95} for the source LMC X-1. Therefore, we consider only open clusters with distances larger than 230~pc.\\
\begin{table*}[t]
\caption{{\bf Testing the models for individual objects} \protect\\
For each object the known distance and hydrogen column density N(H) with references are given. Furthermore are listed for each model: the obtained minimal distance D$_{MIN}$ for the lowest N(H) value, the maximal distance D$_{MAX}$ for the largest N(H) value and the expected N(H) for the known distance. 
The models are the same up to a distance of 230~pc (see text for details).
The first 4 objects are neutron stars with VLBI or HST parallaxes and X-ray--obtained N(H). \small{TY~CrA} lies within the CrA star forming molecular cloud core close to \small{RX J1856.5-37541. HD 18190} is thought to be situated behind \small{MBM~12.} \protect\\
1: \cite{Golden2005}; 2: \cite{Marshall2002}; 3: \cite{Dodson2003}; 4: \cite{Pavlov2001}; 5: \cite{Caraveo1996}; 6: \cite{Jackson2005}; 7:  \cite{Kaplan2002}; 8: our EPIC-pn fit; 9: \citet{Casey1998}; 10:  \cite{Forbrich2006}; 11: \cite{Hobbs1988}; \protect\\
 $^a$ $log($N(Na{\sc{i}}$))=10.6 <11$ \cite{Ferlet1985} formula was doubted by \cite{Welty1994}, see text; $^b$ note the recently refined parallax by \citet{Kerkwijk2006aph} yielding a distance of $161^{+18}_{-14}$~pc; $^c$ mean value; $^d$ lower limit for N(Na{\sc{i}}) was converted to N(H); $out$: needed N(H) is not reached up to the here considered 1~kpc distance
}
\centering
\label{tab:1}       % Give a unique label
\begin {tabular}{ l l c l c | c c c | c c c}
\hline\noalign{\smallskip}
Name & Distance & Ref.  & N(H) & Ref.& \multicolumn{3}{c}{analytical model} & \multicolumn{3}{c}{extinction model}\\[3pt]
 &  &  & & & D$_{MIN}$ & D$_{MAX}$ & N(H) & D$_{MIN}$ & D$_{MAX}$ &N(H)\\[3pt]
% & [pc]  &  & [$10^{20}$ cm$^{-2}$] & & [pc] & [pc]& \small{[$10^{20}$ cm$^{-2}$]} & [pc] & [pc] & \small{[$10^{20}$ cm$^{-2}$]} \\
 & [pc]  &  & \tiny{[$10^{20}$ cm$^{-2}$]} & & [pc] & [pc]& \tiny{[$10^{20}$ cm$^{-2}$]} & [pc] & [pc] & \tiny{[$10^{20}$ cm$^{-2}$]} \\
\tableheadseprule\noalign{\smallskip}
\small{PSR B0656+14} & $288^{+33}_{-27}$ & 1 & $1.73 \pm 0.18$ & 2 & 240 & 260 & 2.68 &230 & 240 & 2.86\\
\small{Vela} & $287^{+59}_{-34}$  & 3 & $3.3 \pm 0.3$ & 4 & 300 & 330 & 2.73 & 860 & 910 & 1.07\\
\small{Geminga} & $157^{+19}_{-17}$ & 5 & $1.76 \pm 0.95 $ & 6 & 220 & 300 &0.09$^a$ & 220 & 240 & 0.09$^a$\\
\small{RX J1856.5-37541}& ${140 \pm 40}$$^b$ & 7 & $ 0.74 \pm 0.10$ & 8 & 130  & 140 & 0.87 & 130 &  140 & 0.87\\
\hline
\small{TY CrA} & $129\pm 11$ & 9 &  $130 \pm 0.3 $$^c$ & 10 & out & out & 0.08 & out & out & 0.08  \\
\small{HD 18190}&  185 & 11 & $>7.21^d$& 11 & 440 & $\cdots$ & 5.05 & 800 &  $\cdots$ & 5.05 \\
\noalign{\smallskip}\hline
\end{tabular}
\vspace{-0.4cm}
\end{table*}
We compared the hydrogen column densities we inferred from the open cluster reddenings with those obtained by the models. 
628 out of the 650 open clusters by \cite{Piskunov2006} lie within the considered data cube with distances larger than 230~pc and smaller than 4.5~kpc. The scatter in the obtained extinction differences increases with distances for both models (see Fig.~\ref{fig:Pisk1}). For the model based on \cite{Hakkila97} the mean deviation is around $9 \cdot 10^{20}$ cm$^{-2}$, the corresponding mean value for the analytical model is $27 \cdot 10^{20}$ cm$^{-2}$. Considering only distances up to 1~kpc these values are $3.3 \cdot 10^{20}$ cm$^{-2}$ and $8.1 \cdot 10^{20}$ cm$^{-2}$ respectively. Interestingly at some galactic longitudes ($l \approx 0^{\circ}$, $120^{\circ}$, and $300^{\circ}$; see Fig.~\ref{fig:Pisk2}) and at the galactic latitude $b \approx 0^{\circ}$ the scatter is more pronounced.
The extinction routine by \cite{Hakkila97} predicts the observed properties of the open clusters better than the analytical model. This can be partly explained by the use of (older) open cluster data (e.g. \citealt{Fitzgeral1968}) in the routine by \cite{Hakkila97}.
The analytical model tends to overpredict the extinction by an order of magnitude, especially at larger distances. 
Due to the convincing advantages of extinction data derived from open clusters, we include them finally as local inhomogenities in our models for distances larger than 230~pc. Again, we take into account that the column density can only increase with distance. If N(H) outside 230~pc is lower than the value at 230~pc in this direction we consequently changed the values based on the \citet{Lall03} to the lower value until an N(H) increasing  homogeneously with distance is reached.\\
Aiming to apply our N(H)-models to X-ray detected neutron stars their performance for well known sources is interesting. We selected four of the rare NSs having at the same time known parallaxes and hydrogen column densities with small error bars obtained from X-ray observations.  
In Table~\ref{tab:1} we present the model-derived distances for the measured N(H) range as well as the model-derived N(H) at the given parallactic distance. While the analytical model gives reasonable results for PSR B0656+14, Vela and RX J1856.5-37541 (RXJ1856 in the following), the extinction model based on \cite{Hakkila97} underestimates the column density towards Vela significantly. As both ISM models are equal up to 230~pc, both give the same minimal distance of 220~pc towards Geminga which deviates from the claimed distance by \cite{Caraveo1996}. Interestingly Walter et al. 2006 (this volume) reported recently  a revised distance of\\ $254^{+111}_{-59}$~pc towards Geminga.   
In the case of RXJ1856 the derived minimal and maximal distances are completely included in the N(H) cube obtained by the \cite{Lall03} measurements. Here, an error of 25~pc applies while an error estimate is difficult for the analytical model at distances larger than 230~pc. The \cite{Hakkila97} model has often a formal mean error of the same order as the obtained N(H) values.

As noted in above it is still debated how good
Na{\sc{i}} D-lines are as tracer for hydrogen. In particular it is unclear whether molecular hydrogen is well traced. 
%In very dense parts of the ISM one could also suspect a worse sampling bjects influencing the column densities of \cite{Lall03} and our N(H) models.
In Table~\ref{tab:1} we include the results obtained for two further test objects situated in or close to molecular
cloud cores -- the star-forming site CrA close to RXJ1856 and MBM12
\citep{Magnani1985}. While towards MBM12 there is an enlarged but still
underestimated amount of hydrogen (see also \citealt{Lall03} for discussion of MBM12), the small CrA cloud is clearly missed. 
Both examples indicate that one has to be cautious in applying the $N(H)$ cube obtained from the \citet{Lall03} measurements in the direction of small molecular clouds.
However, there are only few such density enhancements of molecular hydrogen
in the close Solar neighbourhood. The extinction model as well as the
analytical model are in principle sensitive to molecular hydrogen. At large
distances $>1$~kpc the extinction model may miss molecular clouds due to
their low angular size and bad sampling while the analytical model does
not consider individual clouds.
\vspace{-0.2cm}
\section{Applications of the models}
\label{sec:applic}
\vspace{-0.4cm}
\subsection{Distance estimates}
A possible application is to estimate the distance to neutron stars without parallaxes or other distance measurements. This is the case of the majority of the M7 where 
%up to now only a parallactic distance towards RX~J1856.5-37541 is reported in the literature. 
despite for RXJ1856 only a preliminary parallactic distance of $330^{+170}_{-80}$~pc is reported for RX~J0720.4-3125 by \citealt{Kerkwijk2006aph} (this volume).
We obtain N(H) by considering the best $XSPEC$-fits for blackbody and absorption lines of all currently available XMM EPIC-pn observations reduced with XMM SAS 6.5 for each neutron star. As noted by Haberl 2006 (this volume), in this respect the XMM EPIC-pn  is the best-suited instrument for relative measurements with reasonably small errors ($\approx 0.1 \cdot 10^{20}$ cm$^{-2}$ for all objects). 
The model predictions for these column densities can be found in Table~\ref{tab:M7}.
As noted before, the technical sampling is 10 pc. If a N(H) value lies between two sampling points we indicate this by noting e.g. 23{\bf 5}. However, even errors in the best-studied regions 
($\leq 230$~pc) are around 25~pc , and can be much larger otherwise.
\begin{table}[t]
\vspace{-0.1cm}
\caption{{\bf Distances obtained for the M7} \protect\\
The N(H) was obtained by blackbody and absorption line ($tabs$) $XSPEC$-fits from XMM-$Newton$ EPIC-pn observations (abundances and cross-sections are chosen according to \citealt{Wilms2000}). The number of lines used is indicated in parenthesis. The corresponding model prediction of the distance is given for the analytical model: d$_{ana}$, the extinction model based on \cite{Hakkila97} d$_{ext}$, and the analytical model plus consideration of possible 30\% underprediction by the sodium D-lines: d$_{ana130}$.  See also text for discussion.}
\centering
\label{tab:M7}       % Give a unique label
\begin{tabular}{lc|ccc}
\hline\noalign{\smallskip}
name & N(H) ($\#$lines)& d$_{ana}$ & d$_{ext}$ & d$_{ana130}$ \\[3pt]
 & [$10^{20}$ cm$^{-2}]$ & [pc]& [pc]& [pc]\\
\tableheadseprule\noalign{\smallskip}
RX J1856.5-3754 & 0.7 (0L)& 135 & 135 & 125 \\
%\hline
RX J0420.0-5022 & 1.6 (1L)& 345 & $\cdots$& 325\\
RX J0720.4-3125 & 1.2 (1L)& 270 & 235 & 265 \\
RX J0806.4-4123 & 1.0 (1L)& 250 & 235 & 240\\
RBS 1223 & 4.3 (1L)& $\cdots$& $\cdots$& $\cdots$\\
RX J1605.3+3249 & 2.0 (3L)& 390 & $\cdots$& 325\\
RBS 1774 & 2.4 (1L)& 430 &  $\cdots$& 390\\
\noalign{\smallskip}\hline
\end{tabular}
\vspace{-0.4cm}
\end{table}
In case of the extinction model based on \cite{Hakkila97} one cannot always derive a convincing value due to bad sampling with the (low) extinction not changing over a large scale of distances (e.g. up to 1~kpc). For the analytical model only the high latitude object RBS 1223 is a problem where at 1 kpc one arrives only at $3.2 \cdot 10^{20}$ cm$^{-2}$. The limitations of this large-scale model appear here. 
However, absorption lines in the spectrum also influence N(H). 
Schwope 2006 (this volume) fitted the spectrum of RBS 1223 using recent new observations and obtained N(H)$=3.7 \cdot 10^{20}$~cm$^{-2}$ for one absorption line. This value is comparable to our result in Table~\ref{tab:M7}, taking into account the additional data. However, the fit was not very good and Schwope 2006 obtained better fits with two absorption lines N(H), yielding values ranging from $1.2 \cdot 10^{20}$~cm$^{-2}$ to $1.8 \cdot 10^{20}$~cm$^{-2}$. The lower value would correspond to a distance of 525~pc when using the analytical model.\\
Overall, the distances derived from both models towards two neutron stars are in acceptable agreement: \\for RX~J0720.4-3125 with 250~pc and for RX~J0806.4-4123 with 240~pc. Since both NSs are close to the border of the sodium measurements, an error of $\approx 25$~pc seems reasonable. In general, the values of the analytical model have to be used with caution as local clumpiness of the ISM is not considered by this model at distances $>230$~pc. Considering the open cluster test from above, the mean N(H) deviation of the analytical model up to 500~pc is $ 0.05 \cdot 10^{20}$ cm$^{-2}$; the standard deviation, however, is $ 4.9 \cdot 10^{20}$ cm$^{-2}$.
%$3.4 \cdot 10^{20}$ cm$^{-2}$ and $ 2.2 \cdot 10^{20}$ cm$^{-2}$ respectively.
%bis 750	0.034	0.022	0.056	0.043
%bis 500pc	-0.00047	0.00758	0.049	0.037
\vspace{-0.6cm}
\subsection{Population synthesis}
\begin{figure}
\vspace{-0.2cm}
\centering
  \includegraphics[angle=0,width=0.45\textwidth]{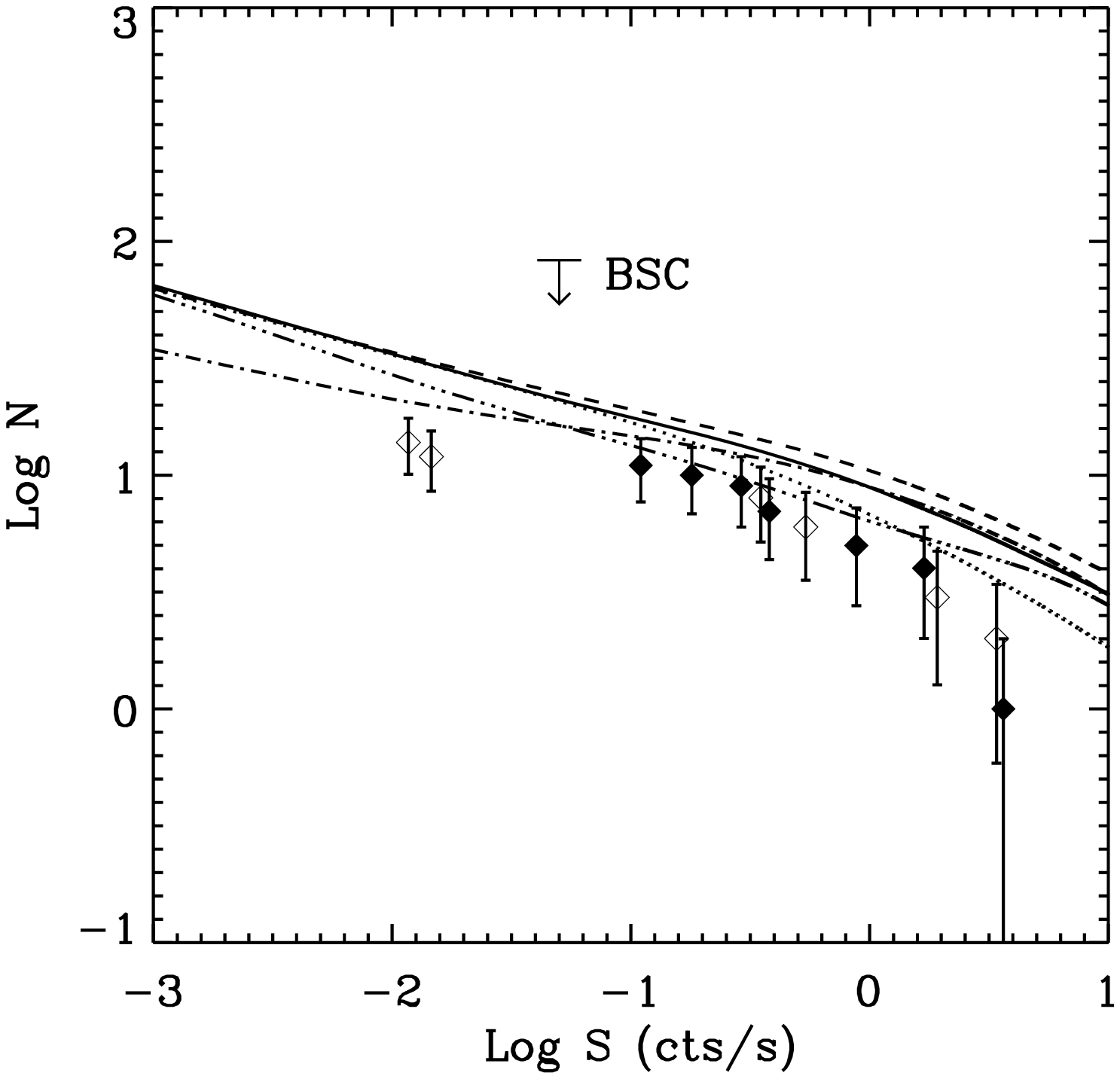}
\caption{{\bf The log$N$--log$S$ curves for different models} \protect\\ 
Each curve shows the expected number of observable neutron stars at ROSAT PSPC countrates.
The solid line represents the curve for the new progenitor distribution but old absorption model; the dotted and  the dashed line are the results of the old simulations by e.g. \citet{Popov03} for Gould Belt radii of $R_{GB}=500$~pc  and $R_{GB}=300$~pc respectively. The only difference to the solid curve is the progenitor distribution.
The results of taking into account the new progenitor distribution and additionnally the new absorption models are shown by the dashed-dotted line for the refined analytical ISM model and the dot-dot-dot-dashed line for the extinction model based on \cite{Hakkila97}. Furthermore the measurement points of young NSs with thermal X-ray emission are plotted as in \cite{Popov03}. For discussion see text.
}
\label{fig:logN}
       % Give a unique label
\vspace{-0.8cm}
\end{figure}
The main motivation for us to take into account better absorption treatment is an improved population synthesis model based on that of \cite{Popov2000,Popov03} and \cite{Popov05}. A detailed discussion of the improvements is beyond the scope of this article and will be provided elsewhere (Popov, Posselt, \emph{in preparation}). Therefore, we summarize here only the new developments and concentrate on illustrating the importance of the interstellar absorption.\\
While \cite{Popov03,Popov05} modeled the neutron star progenitor distribution as coming from infinitely thin disks, either from the galactic plane or the Gould Belt, the new initial distribution is more realistic. Up to the Hipparcos limit of 400~pc \citep{Hip1997} the known B2-O8 stars are considered individually as well as their affiliation to an OB association \citep{Zeeuw99} resulting in birth properties depending on the age of the OB-association. The Gould Belt birth rate is applied here. For distances above 400~pc, the galactic disk is considered for few neutron stars randomly as well as 36 associations \citep{Blaha1989,Melnik95} for most of the neutron stars. Due to the unknown ages of the associations, the birth probability is set proportional to the number of association members.
%Furter improvements inlude a newer neutron star birth mass spectrum \citep{Heger2004} with a changed relation between progenitor and neutron star mass for progenitors with
%initial masses $>12\, M_{\odot}$, and the application of the cooling curve set discussed as .\\
In Fig.~\ref{fig:logN}, the log$N$-log$S$ curves are shown for the old thin-disk Gould Belt models with either 300~pc or 500~pc radius, the new progenitor distribution as well as for the the new progenitor distribution considering both absorption models. All simulations were done for the same mass spectrum  as in \cite{Popov2006}, the cooling curve set labeled Model~I in  the same paper, the same abundances and cross-sections (in this case those of \cite{MM83} for comparison) and identical remaining parameters.
The results for the new progenitor distribution lie in-between those of the two Gould Belt sizes, a little apart from the measured integrated number of objects. This is also true for the curve considering the absorption by the analytical model. At the bright end the differences introduced by the analytical absorption model are only very small. Towards lower countrates, the curve lies significantly below the one without the refined 3D ISM model.  Considering the absorption based on the extinction model by \cite{Hakkila97}
gives another picture. At the bright end the curve is much lower than that without the refined 3D ISM model, while the difference becomes smaller towards low countrates. We note here again that both absorption models are the same up to 230~pc, based on the measurements by \cite{Lall03}. Both also consider the open cluster data, however slightly differently as the column density is only allowed to increase. The curve of the analytical model shows best the influence of an improved ISM distribution model compared to the old absorption treatment (e.g. \citealt{Popov03}) where the analytical model without the data by \cite{Lall03} and \cite{Piskunov2006} was used). 
One might have expected that differences should be largest at low distances and hence for bright sources. However, averaging over the small Local Bubble gives apparently the same result as the simpler description before. 
The influence of the better model for the Solar vicinity is apparent in the expected location likelihoods for neutron stars. Going to larger distances the additive column densities are especially large around $l \approx 300^{\circ}$ to $l \approx 60^{\circ}$ in the galactic plane where many objects could be expected. Thus, the number of predicted neutron stars is reduced. The N(H) model based on the extinction study by \cite{Hakkila97} shows on average much higher N(H) already at 240~pc compared to the analytical model. Given the nearly rectangular shape of the regions with relatively high values in the $l,b$ plane these seem to be caused at least partly by bad sampling and the necessary statistical treatment. The higher N(H) results in less observable sources bringing the corresponding log$N$-log$S$ much closer to the actually measured number of sources at the bright end.\\
%The predicted number of sources per square degrees is shown in fig~\ref{fig:1} for the analytical model as an example for comparison with Fig.~6 of \cite{Popov05}. 
%\begin{figure}
%\centering
%% For example, with the graphicx package use
%  \includegraphics{example.eps}
%\end{figure}
We made similar test plots for the expected neutron star number per square degree like the one presented in Fig.~6 by \cite{Popov05}. We summarize here only the main results, for detailed discussion see Popov, Posselt 2006 (\emph{in prep.}).
Contrary to the old model the predicted neutron star number is enhanced in regions including the nearby Gould Belt OB associations (e.g. Upp Sco) with the new progenitor distribution. Sources are not anymore homogeneously distributed at all longitudes along the planes of the galactic disk or the Gould Belt.
When it comes to consider the 3D ISM models, the influence of the absorption is on much smaller scales. Less sources would be expected at higher latitudes, and towards some small regions the high extinction hinders the possible detection of neutron stars. Differences between the two absorption models are pronounced only in small regions which to discuss in detail is beyond our intention here. 
\newpage
\vspace{-0.4cm}
\section{Conclusions}
An understanding of the interstellar absorption is crucial for the 
interpretation of low-resolution soft X-ray spectra. Taking into account the 3D distribution of the ISM is important to determine the absorption of known sources as well as for searches of new soft X-ray sources towards a particular direction in the sky. We presented two simple models for 3D ISM distributions together with possible applications: distance estimation and the population synthesis for young isolated thermal, thus soft X-rays emitting neutron stars.
Both models are large scale models and have to be used with caution especially considering distance estimations. Concerning population synthesis the two models influence the log$N$--log$S$ curve discriminatively, while the differences for the predicted spatial distribution of observable neutron stars are only small. 


%\section{aaa}
%\label{sec:1}
%%and \cite{Ref1}
%\subsection{Subsection title}
%\label{sec:2}
%as required. Don't forget to give each section
%%and subsection a unique label (see Sect.~\ref{sec:1}).
%\paragraph{Paragraph headings} Use paragraph headings as needed.
%\begin{equation}
%\end{equation}

%\begin{figure}
%\centering
%% Use the relevant command to insert your figure file.
%% For example, with the graphicx package use
%  \includegraphics{example.eps}
%\end{figure}
%%
%% For two-column wide figures use
%\begin{figure*}
%\centering
%% Use the relevant command to insert your figure file.
%% For example, with the graphicx package use
%  \includegraphics[width=0.75\textwidth]{example.eps}
%\end{figure*}
%

%\begin{table}[t]
%\centering
%\begin{tabular}{lll}
%\hline\noalign{\smallskip}
%first & second & third  \\[3pt]
%\tableheadseprule\noalign{\smallskip}
%\end{tabular}
%\end{table}

\vspace{-0.4cm}
\begin{acknowledgements}
We would like to thank R. Lallement providing us with the sodium density cube and for discussions as well as A. Piskunov for answering our questions regarding the open cluster study.\\
S.B.Popov also acknowledges the grants RFBR 04-02-16720 and 06-02-26586.
\vspace{-0.3cm}
\end{acknowledgements}
%\begin{tiny}
% BibTeX users please use
%\bibliographystyle{spmpsci}
%\bibliographystyle{aa}
%\bibliography{Drbib}   % name your BibTeX data base
%\end{tiny}


%\begin{thebibliography}{3}
%%
%% and use \bibitem to create references. Consult the Instructions
%% for authors for reference list style.
%\bibitem{Ref1}
%Author, I.: Article title. Journal Title-Abbreviated {\bf Vol}, pp--pp (year)
%\bibitem{Ref2}
%Author, I., Smith, J.: Book Title. Publisher, Place (year)
%% Format for proceedings
%\bibitem{Ref3}
%Author, I., Smith, J.: Paper title. In: Editor, A. (ed.) Proceedings
%Title, Location, Date, pages. Publisher, Place (year)
%\end{thebibliography}

\end{document}